# Research on Mechanical Properties and Deformation-Fracture Energy Consumption Characteristics of Plateau Frozen Rocks

Hongbing Yu [1,2], Jiyu Wang [3], Xiaojun Zhang [1*], Mingsheng Zhao [2,4]

[1] College of Architecture and Civil Engineering, Beijing University of Technology, Beijing, 100124, China; whutyhb@hotmail.com;

[2] Hunan Lianshao Construction Endineering Group Co.,LTD, Changsha Hunan, 410000, China; whutyhb@hotmail.com;

[3] Poly Explosive Hami Co., Ltd., Hami, Xinjiang, 839000, China; qawangjiyu@163.com;

[4] Hongda Blasting Engineering Group Co., Ltd., Guangzhou Guangdong, 510000, China, ms1027_zhao@hotmail.com;

*Correspondence: zhangxj@bjut.edu.cn; Tel.: +86 18210136152

**Abstract**

The exploitation of mineral resources in plateau regions is confronted with critical challenges including low blasting efficiency, excessive energy consumption, and compromised operational safety when dealing with low-temperature water-bearing frozen rock masses. This study systematically investigates the dynamic-static mechanical properties, deformation-fracture behaviors, and energy consumption characteristics of plateau frozen sandstone under the coupled effects of temperature (-20 °C to 20 °C) and moisture content (5%-15%). The research methodology integrates field sampling, low-pressure low-temperature simulation tests, graded impact loading tests, and numerical inversion analysis. Results demonstrate that freezing significantly enhances the dynamic strength and brittleness of saturated sandstone. The pore structure undergoes substantial evolution with decreasing temperature, with the porosity increasing by 63.15% at -20 °C. Based on PFC3D microscopic simulations, the mechanism of frost heave damage and the regulatory effect of water-ice phase transition on rock mechanical behaviors are elucidated. A quantitative analysis method for energy dissipation is proposed, revealing that the energy absorption increment of frozen rocks is higher than that of room-temperature samples. The findings provide a theoretical basis and technical support for optimizing blasting parameters, realizing directional energy release, and promoting green construction of frozen rock masses in high-altitude areas.

## 1. Introduction
### 1.1 Research Background

Plateau regions serve as the core supply base for global strategic mineral resources, and their development progress directly impacts the international resource security pattern. The unique climatic and geological conditions in these areas lead to the water-rich freezing coupling characteristics of rock masses, which significantly alter their dynamic mechanical responses and wave impedance properties. This further gives rise to key technical bottlenecks such as intensified energy dissipation during blasting and uneven rock fragmentation. These issues not only result in non-linear increases in construction costs but also lead to ineffective control of blasting hazards (e.g., flying rocks, vibration), severely restricting the coordinated achievement of resource development efficiency and environmental sustainability goals. To address this international technical challenge, it is imperative to establish a theoretical system for the blasting effect of water-bearing frozen rock masses, develop innovative blasting technologies based on energy-oriented regulation, and provide scientific support for safeguarding the security of global resource supply chains and the sustainable development of ecologically fragile high-altitude regions.

**1.2 Research Objectives**

Aiming at the technical bottlenecks of frozen rock mass blasting in high-altitude areas, this study systematically analyzes the dynamic mechanical responses and fracture energy consumption characteristics of such rock masses, and constructs a theoretical system for blasting optimization based on energy-oriented regulation. Specifically, the research objectives include: (1) Quantifying the attenuation law of stress wave propagation and the fractal expansion characteristics of cracks under frozen conditions, so as to provide dynamic basis for the optimization of charge structure and hole network parameter design; (2) Establishing an adaptive blasting parameter model considering the coupled effects of moisture content and temperature, to realize the intelligent matching of seasonal construction parameters; (3) Forming the ultimate technological path of "efficient energy conversion-precise hazard control-minimized ecological disturbance", which can significantly improve the economic efficiency and environmental friendliness of blasting operations. The research is expected to provide core theoretical support and technical paradigms for resource development in high-altitude cold environments.

**1.3 Research Status**

As a natural geological material, rocks contain various internal defects such as micropores and joints, which provide storage space for water. The mechanical properties of rocks are significantly affected by temperature changes and moisture content [1-7]. Existing studies have shown that the mechanical properties of frozen rocks exhibit distinct temperature-moisture dual-control characteristics: in the low saturation range (<60%), their strength is mainly dominated by the thickness of the unfrozen water film; under high saturation conditions (>90%), the pore ice content becomes the key controlling factor, and the frost heave damage effect increases exponentially as saturation approaches 100% [8]. Temperature gradient experiments have revealed the existence of a dynamic strength extremum point near -30 °C [9-10], with an ice crystal strengthening dominant zone (-30~0 °C) and a frost heave damage dominant zone (<-30 °C) on either side of this critical temperature. Notably, moisture content gradient tests have indicated a non-linear relationship between strength growth rate and initial moisture content [11], where the inflection point corresponds to the critical saturation at which the pore water-ice phase transition is completed. In addition, frozen soft rocks under unloading paths exhibit significant expansion anisotropy [12], which provides a new mechanical criterion for the stability evaluation of deep frozen rock engineering.

**1.4 Research Gaps**

Current research on the blasting of water-bearing frozen rock masses in plateaus still faces several key scientific issues that need to be addressed. In terms of experimental simulation, existing studies fail to effectively reproduce the real response of rock masses under the coupled effects of multiple fields such as low pressure, moisture content (saturated, dry), and temperature difference in plateau environments. At the level of failure mechanism, the cross-scale correlation between crack evolution under dynamic loading conditions and ice-rock phase transition damage has not yet been established. In the dimension of energy regulation, there is a lack of quantitative characterization of the dispersion and attenuation characteristics of blasting stress waves in frozen rock masses. In the field of engineering application, current blasting designs still rely on empirical correction coefficients. In terms of safety control, environmental disturbance prediction remains in the qualitative evaluation stage, and there is an urgent need to develop a comprehensive green blasting evaluation system based on energy flow density distribution.

**1.5 Research Scope and Content**

This study addresses the problems of large discrepancies between experimental conditions and actual environments, insufficient understanding of dynamic mechanisms, lack of quantitative description of energy dissipation characteristics, and absence of targeted optimization models for blasting parameters in high-altitude water-bearing frozen rock mass blasting. A research system integrating field sampling, low-pressure low-temperature simulation tests, graded impact loading tests, and numerical inversion analysis is constructed to reveal the dynamic strength, brittleness, and energy distribution laws of rock masses under the coupled effects of temperature and moisture content. A blasting dynamic constitutive model and parameter optimization method suitable for high-altitude frozen rock masses are established. The research scope covers sandstone with an altitude of 3000-4500 meters, a moisture content of 5%-15%, and a low-temperature range of -5 °C to -20 °C. Results show that optimized blasting design based on dynamic parameter correction can significantly reduce explosive consumption, control flying rocks and vibration, and balance construction efficiency and environmental protection.

**2. Analysis of Dynamic-Static Mechanical Failure and Deformation Characteristics of Frozen Rocks**

**2.1 Sample Preparation and Experimental Scheme**

In this study, primary sandstone collected from a mine in Tibet (altitude: 4569-5118 m) was used to fabricate standard specimens with dimensions of Φ100mm×50mm (for static load tests) and Φ25mm×50mm (for dynamic load tests). The specimens were screened by longitudinal wave velocity (measured via ultrasonic testing, with a screening threshold of ±5% deviation from the average velocity) to ensure uniformity, and then subjected to quasi-static tests using the MTS 815 Rock Mechanics Test System and dynamic impact tests using the SHPB (Split Hopkinson Pressure Bar) system (diameter: 50 mm), respectively.

Based on the extreme climatic characteristics of the mining area (annual temperature range: -19.4~17.5 °C), three temperature gradients were set: 20 °C (room temperature), -10 °C, and -20 °C. The saturation treatment adopted the vacuum pumping method (vacuum degree: -0.095 MPa, saturation time: 72 hours) to ensure full water absorption of specimens, and the freezing treatment was conducted in a high-precision low-temperature chamber (temperature control accuracy: ±0.5 °C, freezing time: 48 hours) to simulate actual freeze-thaw conditions.

Specimen pretreatment included two key steps: (1) Low-temperature strain gauges (BE120-5AA, working temperature range: -50~150 °C) were pasted on the axial and radial surfaces of specimens, and a 2mm-thick silicone rubber insulation layer (thermal conductivity: 0.18 W/(m·K)) was coated to prevent frost formation and ensure strain measurement accuracy; (2) A dynamic strain acquisition system (DH5922, sampling rate: 1 MHz) was synchronized with the low-temperature chamber and loading equipment to realize real-time acquisition of temperature, strain, and load data.

The key physical and mechanical parameters of the rock samples (measured via standard rock mechanics tests) are presented in Table 1, and the experimental system configuration (including sampling, specimen preparation, and testing equipment) is detailed in Figure 1.

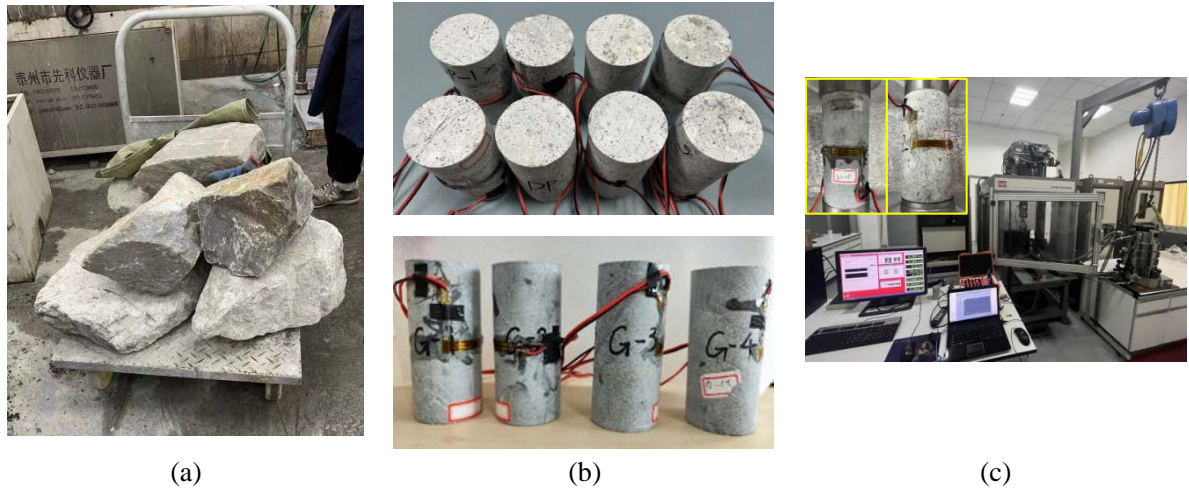

| (a) | (b) | (c) |

Figure 1 Preparation of specimens and MTS system

*Note: (a) Field sampling of sandstone: Samples were collected from the shallow surface (5-8 m depth) of the mine to avoid the influence of deep in-situ stress; (b) Specimen preparation process: Using a diamond saw to cut raw rock into standard cylinders, followed by grinding to ensure that the end face flatness error is less than 0.02 mm and the parallelism error is less than 0.05 mm; (c) MTS 815 Rock Mechanics Test System: Equipped with a low-temperature environmental chamber, capable of applying axial loads up to 2600 kN and simulating temperatures as low as -60 °C.*

Table 1 Physical and mechanical parameters of sandstone (average values of 5 parallel specimens)

| Rock sample type | Density /kg·m$^{-3}$ (measured via drainage method) | Porosity /% (measured via helium porosimetry) | Longitudinal wave velocity /km·s$^{-1}$ (measured via ultrasonic testing) | Elastic modulus /GPa (calculated via static uniaxial compression) | Poisson's ratio (calculated via static uniaxial compression) |
|---|---|---|---|---|---|
| Sandstone | 2.71±0.03 | 8.59±0.21 | 4.74±0.08 | 8.40±0.35 | 0.20±0.02 |

**2.2 Static Stress-Strain Characteristics**

Static uniaxial compression tests were conducted at a loading rate of 0.05 mm/min (to avoid dynamic effects), and the test results (Figures 2 and 3) indicate that frozen sandstone specimens exhibit a typical tensile-shear composite failure mode—characterized by primary shear cracks accompanied by secondary tensile cracks on the specimen surface. Within the temperature range of -20~20 °C, the failure mode does not undergo significant changes, suggesting that the freezing effect has a limited impact on the macroscopic failure mechanism of low-porosity sandstone (n<8%). This is because the dense mineral skeleton of low-porosity sandstone restricts the expansion space of pore ice, reducing the degree of frost heave-induced structural damage.

Measurements of elastic parameters (calculated from the linear elastic segment of stress-strain curves, with a strain range of 0.05%~0.15%) yield a Poisson's ratio (μ) of 0.20±0.02 and an elastic modulus (E) of 4.0±0.2 GPa. Comparison of stress-strain curves between saturated (SS) and dry (SD) specimens (Figure 3) reveals three key phenomena: (1) The saturation effect reduces the uniaxial compressive strength of room-temperature specimens by 23.5% (average value, from 52.3 MPa to 40.0 MPa), which is attributed to the softening effect of water on mineral interfaces—water infiltrates into grain boundaries, reducing the friction coefficient and bonding strength between minerals; (2) The strengthening effect of freezing on saturated specimens (Δσc=18.7 MPa, from 40.0 MPa at 20 °C to 58.7 MPa at -20 °C) is significantly higher than that on dry specimens (Δσc=5.3 MPa, from 52.3 MPa at 20 °C to 57.6 MPa at -20 °C), as pore ice in saturated specimens forms a "cementing phase" that fills microcracks and enhances the integrity of the rock skeleton; (3) At -20 °C, all specimens exhibit stress drops in the elastic stage (strain energy release U>0.85 kJ/m³, calculated via the integral of stress-strain curves), confirming the brittle transition effect induced by ice crystal bonding. This phenomenon originates from the competitive advantage mechanism of ice phase strengthening (elastic modulus of ice, $K_{ice}$=9 GPa) over the water softening effect in low-porosity sandstone—ice crystal bonding offsets the interface weakening caused by water and further enhances the rigidity of the rock skeleton.

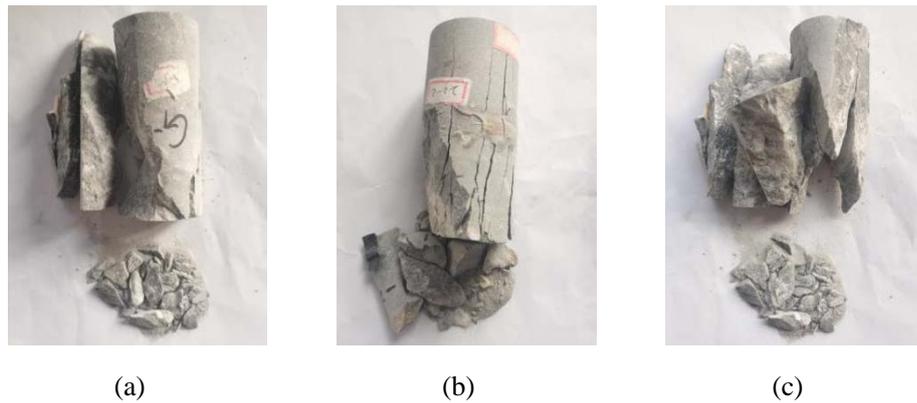

(a)             (b)             (c)

Figure 2: Fracture morphology of the specimen

*Note: (a) 20 °C (saturated): Mainly powdery failure, with a large number of fine fragments; (b) -10 °C (saturated): Blocky-powdery mixed failure, with obvious shear planes; (c) -20 °C (saturated): Typical blocky failure, with a single main shear crack and few secondary cracks; (d) 20 °C (dry): Blocky failure, with more secondary cracks than saturated specimens at the same temperature; (e) -10 °C (dry): Blocky failure, similar to saturated specimens but with smaller fragment sizes; (f) -20 °C (dry): Blocky failure, with the largest fragment size among all groups.*

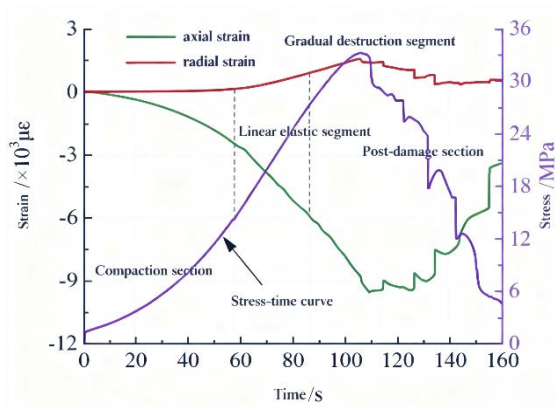 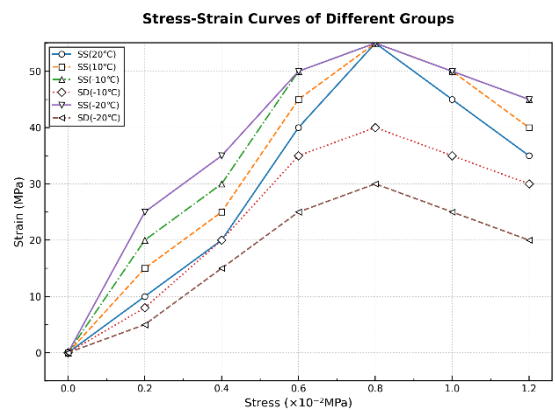

Figure 3 Deformation and strength characteristics during the loading process of the specimen

Figure 4 Static stress-strain curves at different temperatures

*Note: (a) Deformation and strength characteristics during loading: The curve is divided into four*

*stages—compaction stage (strain <0.1%), linear elastic stage (0.1%<strain <0.3%), plastic yield stage (0.3%<strain <0.4%), and post-failure stage (strain >0.4%); the stress-time curve shows a linear increase in the elastic stage and a sudden drop after peak stress, reflecting brittle failure characteristics. (b) Static stress-strain curves at different temperatures: SS represents saturated specimens, SD represents dry specimens; the peak stress of SS specimens increases significantly with decreasing temperature, while the peak stress of SD specimens changes slightly; the peak strain of all specimens decreases with decreasing temperature, indicating increased brittleness.*

**2.3 Influence of Temperature on Sandstone Porosity**

A MesoMR12-060H-I nuclear magnetic resonance (NMR)(Figure 5) imaging system (resonance frequency: 12 MHz, magnetic field strength: 0.3 T) was used to determine the pore structure evolution characteristics of sandstone under frozen conditions (20 °C, -10 °C, and -20 °C). The test adopted the CPMG (Carr-Purcell-Meiboom-Gill) pulse sequence, with an echo time of 0.2 ms and a number of echoes of 10,000, to measure the transverse relaxation time (T2) of pore water—where T2 is positively correlated with pore size (larger pores correspond to longer T2 values).

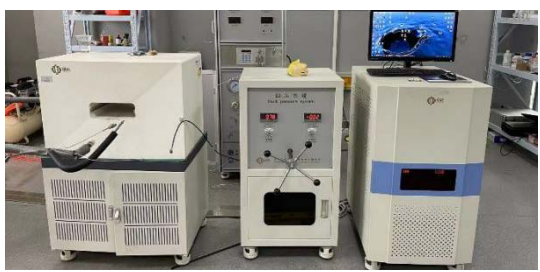

Figure 5 MesoMR12-060H-I Magnetic Resonance Imaging Analysis System

T2 spectrum analysis shows that freezing significantly alters the pore distribution characteristics of sandstone, specifically manifested as two key trends: (1) A 37.2% increase in the proportion of micropores (T2<10 ms, corresponding to pore size <100 nm) at -20 °C compared to 20 °C, which is attributed to the shrinkage of mineral particles caused by low temperature, leading to the generation of new microcracks; (2) A non-linear decrease in effective porosity (calculated via the integral of T2 spectrum area, calibrated with helium porosimetry) with decreasing temperature—from 8.59% at 20 °C to 6.15% at -20 °C, a reduction of 28.6%—due to the expansion of pore ice (volume expansion rate of water freezing: ~9%), which fills small pores and blocks pore throats.

Table 2 further reveals that decreasing freezing temperature substantially modifies the pore structure distribution of sandstone through three mechanisms: (1) The relaxation time spectrum exhibits a three-peak structure (Peak 1: T2<10 ms, micropores; Peak 2: 10 ms<T2<100 ms, mesopores; Peak 3: T2>100 ms, macropores). As temperature decreases from 20 °C to -20 °C, the amplitude of each peak increases and shifts to the right (the area of peaks 1 and 2 increases by 37.2%), confirming that the melting-shrinkage effect induced by ice phase transition (ice melts and shrinks when temperature rises, but the reverse process—freezing expansion—enlarges pore size during cooling) enlarges the pore size of micropores (<100 nm) and mesopores (100-1000 nm); (2) Pore composition undergoes significant reorganization: the proportion of small pores (T2<10 ms) decreases from 92.63% (room temperature) to 89.84% (-20 °C), the proportion of medium pores (10-100 ms) increases from 7.16% to 9.82%, while large pores (>100 ms) remain stable (<0.5%)—this is because large pores have sufficient space for ice expansion, avoiding further pore enlargement; (3) Total pore volume (characterized by T2 spectrum area) expands significantly, with the spectral area change rate surging from 22.21% at -10 ℃ to 63.15% at -20 ℃ (calculated relative to room temperature). This indicates that low-temperature freezing exerts a significant pore modification effect on low-permeability sandstone, providing a microstructural explanation for the strength strengthening mechanism of frozen sandstone—pore ice filling and pore size optimization enhance the integrity of the rock skeleton, thereby improving mechanical strength.

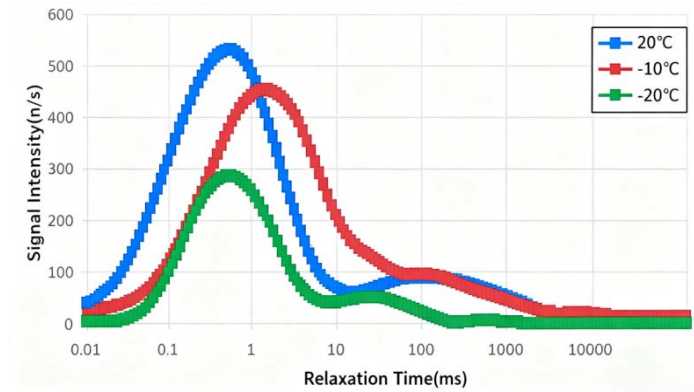

Figure 6 Signal intensity diagram of sandstone relaxation process at different temperatures

*Note: (a) 20 ℃: The spectrum is dominated by Peak 1 (micropores), with a small proportion of Peak 2 (mesopores) and almost no Peak 3 (macropores); (b) -10 ℃: The amplitude of Peak 1 and Peak 2 increases, and the peak position shifts to the right (pore size increases); (c) -20 ℃: The amplitude of all peaks further increases, and the proportion of Peak 2 increases significantly, indicating that more micropores are transformed into mesopores due to freezing expansion.*

Table 2 Statistical parameters of T2 spectra of frozen sandstone at different temperatures (unit: %, average values of 3 parallel specimens)

| number | temperature /℃ | Proportion of each peak area /% | | | Spectral area | Average spectral area | Mean change rate of spectral area /% |
| --- | --- | --- | --- | --- | --- | --- | --- |
| | | Peak 1 (T2<10 ms) | Peak 2 (10 ms<T2<100 ms) | Peak 3 (T2>100 ms) | | | |
| A1 | 20 | 92.63±0.35 | 7.01±0.22 | 0.24±0.05 | 14250±320 | 14683±280 | 0±0.5 |
| A2 | 20 | 92.46±0.28 | 7.13±0.18 | 0.20±0.03 | 15240±280 | | |
| A3 | 20 | 92.50±0.31 | 7.12±0.20 | 0.17±0.04 | 14561±300 | | |
| A4 | -10 | 91.49±0.25 | 8.05±0.15 | 0.25±0.03 | 17574±250 | 17944±220 | 22.21±1.2 |
| A5 | -10 | 91.30±0.22 | 8.22±0.12 | 0.27±0.04 | 18400±230 | | |
| A6 | -10 | 91.36±0.24 | 8.14±0.14 | 0.30±0.03 | 17850±240 | | |
| A7 | -20 | 90.08±0.21 | 9.39±0.11 | 0.33±0.02 | 24250±210 | 23956±190 | 63.15±1.5 |
| A8 | -20 | 89.44±0.19 | 10.01±0.10 | 0.34±0.03 | 23830±200 | | |
| A9 | -20 | 89.69±0.20 | 9.75±0.12 | 0.35±0.02 | 23760±210 | | |

## 2.4 Microscopic Characteristics of Sandstone Frost Heave Process

### 2.4.1 Construction of Rock Freezing Model

A linear parallel bonding model (LPBM) in PFC3D (Particle Flow Code 3D) was adopted for discrete element numerical simulations. The model simplifies saturated rock masses into a two-phase medium system consisting of mineral skeleton particles (simulating quartz, feldspar, etc.) and pore water particles (simulating pore water), based on the actual mineral composition (quartz: 65%, feldspar: 25%, clay minerals: 10%) and pore structure of the sandstone [13].

#### 2.4.1.1 Contact Pair Design

Based on contact mechanics theory, three types of contact pairs were established to simulate the interactions between different phases:

- **Rock-Rock contact**: Simulates the bonding between mineral particles, using the parallel bonding model with high stiffness and strength to reflect the rigid connection of the rock skeleton;
- **Rock-Water contact**: Simulates the interaction between mineral particles and pore water, using the linear contact model with low stiffness to reflect the weak bonding between water and rock;
- **Water-Water contact**: Simulates the cohesion between pore water molecules, using the linear contact model with moderate stiffness to reflect the surface tension of water.

The ratio of the number of rock particles to water particles was determined by the porosity (n) of the sandstone, as expressed in Equation (1):

$$n = \frac{V_w}{V_{\text{total}}} = \frac{N_w \cdot V_{w,\text{single}}}{N_r \cdot V_{r,\text{single}} + N_w \cdot V_{w,\text{single}}} \tag{1}$$

Where $V_w$ is the total volume of water particles, $V_{\text{total}}$ is the total volume of the model, $N_w$ and $N_r$ are the number of water particles and rock particles, respectively, and $V_{w,\text{single}}$ and $V_{r,\text{single}}$ are the average volume of a single water particle and rock particle, respectively. For the sandstone in this study (porosity n=8.59%), the number of water particles and rock particles was calculated as 8210 and 56428, respectively (Table 3).

#### 2.4.1.2 Model Resolution Calculation

The model resolution (RES) is defined as the ratio of the model size to the particle size range, which determines the accuracy of the simulation. A higher RES value indicates a more uniform particle size distribution and higher simulation accuracy. The RES was calculated as expressed in Equation (2):

$$RES = \frac{R_{model}}{R_{\max} - R_{\min}} \tag{2}$$

Where $R_{model}$ is the radius of the cylindrical model (25 mm, consistent with the dynamic test specimens), $R_{\max}$ is the maximum particle radius, and $R_{\min}$ is the minimum particle radius. To ensure RES>5 (a common threshold for discrete element models), the maximum and minimum particle radii were set to 1.2 mm and 1.0 mm (rock particles) and 0.95 mm and 0.8 mm (water particles), respectively, resulting in $RES$=25/(1.2-1.0)=125>5, which meets the accuracy requirements.

The final particle parameters are presented in Table 3.

Table 3 Particle parameters of freeze-thaw sandstone in PFC3D model

| Particle Type | density kg/m³ (consistent with experimental measurements) | $R_{min}$ /mm(consistent with experimental measurements) | $R_{max}$ /mm(consistent with experimental measurements) | Number of particles (consistent with experimental measurements) |
|---|---|---|---|---|
| Saturated sandstone water particles | 960 | 0.8 | 0.95 | 8210 |
| Saturated sandstone rock particles | 2600 | 1.0 | 1.2 | 56428 |
| Dry sandstone rock particles | 2600 | 1.0 | 1.2 | 64638 |

### 2.4.1.3 Temperature Field Simulation

The built-in thermal analysis module of PFC3D was used to simulate the heterogeneous temperature field of the specimen. The module calculates the heat transfer between particles based on Fourier's law of heat conduction, as expressed in Equation (3):

$$Q = -k \cdot A \cdot \frac{\Delta T}{\Delta x} \tag{3}$$

where $Q$ is the heat flux, $k$ is the thermal conductivity of the particle material, $A$ is the contact area between particles, $\Delta T$ is the temperature difference between particles, and $\Delta x$ is the distance between particle centers.

Cyclic temperature boundaries were set on the top and bottom surfaces of the cylindrical model (to simulate the gradual cooling process in the low-temperature chamber): the temperature was decreased from 20 °C to the target temperature (-10 °C or -20 °C) at a rate of 1 °C/min, and then maintained for 48 hours (consistent with the experimental freezing time) to ensure uniform temperature distribution inside the model. The real-time temperature at different internal positions (e.g., center, surface) of the model was calculated, and the temperature difference between the center and surface was controlled to be less than 0.5 °C, ensuring consistency with the experimental conditions.

The final numerical model (including particle distribution and contact pairs) is shown in Figure 6.

### 2.4.2 Frost Heave Deformation Calculation

Rock freeze-thaw damage is essentially caused by the volume expansion of pore water during freezing, which exerts frost heave force on the rock skeleton. In the PFC3D model, the frost heave effect was simulated by the thermal expansion of water particles—when the temperature decreases, water particles expand, and the expansion force is transmitted to the rock skeleton through contact pairs, leading to the generation and propagation of cracks.

### 2.4.2.1 Particle Expansion Formula

The radius increment $\Delta R$ of water particles due to thermal expansion was calculated based on the linear thermal expansion theory, as expressed in Equation (4):

$$\Delta R = \alpha \cdot R_0 \cdot \Delta T \tag{4}$$

Where $\alpha$ is the linear thermal expansion coefficient of water/ice, $R_0$ is the initial radius of the particle, and $\Delta T$ is the temperature increment (negative for cooling). For water, the linear thermal expansion coefficient changes with phase state: $\alpha_{water}$=1.769×10⁻⁴/°C (above 0 °C), $\alpha_{ice}$ =2.079×10⁻⁴/°C (below 0 °C) [13]. This phase-dependent coefficient ensures that the expansion of water particles during freezing (0 °C→-20 °C) is larger than that during cooling in the liquid state, reflecting the actual volume expansion of water freezing.

**2.4.2.2 Bond Thermal Expansion Correction**

To simulate the thermal expansion of the bonding between particles (which affects the contact force), the parallel bonding model was modified to consider the thermal expansion of the bond material. Assuming that only the normal component of the bond force is affected by temperature changes (since the model is under uniaxial loading), the normal force increment ($\Delta F_n$) caused by bond expansion was calculated as expressed in Equation (5):

$$\Delta F_n = -k_n \cdot A \cdot \Delta U_n = -k_n \cdot A \cdot (\alpha_b \cdot L_0 \cdot \Delta T) \tag{5}$$

Where $k_n$ is the normal stiffness of the bond, $A$ is the cross-sectional area of the bond (calculated as $A = \pi \cdot (R_r + R_w)^2$, where $R_r$ and $R_w$ are the radii of the two contacting particles), $\Delta U_n$ is the normal displacement of the bond due to thermal expansion, $\alpha_b$ is the linear thermal expansion coefficient of the bond material, $L_0$ is the initial length of the bond, and $\Delta T$ is the temperature increment. The negative sign indicates that the bond force is compressive when the bond expands (temperature decreases).

**2.4.2.3 Bond Strength Setting**

To ensure that the frost heave damage is mainly caused by the expansion of water particles (rather than the fracture of water-water or rock-water bonds), the bonding strength between water particles and between rock and water particles was set to a sufficiently high value (Table 4): the normal and shear bond strengths of water-water bonds and rock-water bonds were 60 MPa, which is much higher than the frost heave force (calculated to be ~10 MPa) generated during freezing. In contrast, the bond strength of rock-rock bonds was set to 40 MPa (consistent with the experimental strength of the sandstone), ensuring that when the frost heave force exceeds the bond strength, rock-rock bonds fracture, simulating the generation of cracks.

**2.4.3 Model Parameters**

The microscopic parameters of the model (e.g., stiffness, strength, thermal conductivity) were calibrated by matching the simulation results with the experimental results (e.g., uniaxial compressive strength, elastic modulus). The calibration process was as follows: (1) Initialize the microscopic parameters based on the material properties of the sandstone (e.g., elastic modulus of rock particles = 9 GPa, consistent with the experimental elastic modulus); (2) Conduct uniaxial compression simulations with different parameter combinations; (3) Compare the simulation results (peak strength, elastic modulus) with the experimental results; (4) Adjust the parameters iteratively until the relative error between simulation and experiment is less than 5%.

The final calibrated microscopic parameters are presented in Tables 4-6.

Table 4 Microscopic parameters of unfrozen and thawed saturated sandstone (calibrated via experimental matching)

| Bond type | $E_m$/GPa (bond elastic modulus) | K (stiffness ratio, ($k_s/k_n$)) | ν (Poisson's ratio of bond) | $P_{be}$/GPa (bond effective modulus) | $P_{bk}$ (bond stiffness ratio) | $P_{bt}$/MPa (bond tensile strength) | $P_{bc}$/MPa (bond compressive strength) | $P_{bfa}$/(°C) (bond friction angle) |
|---|---|---|---|---|---|---|---|---|

| Bond type | $E_m$/GPa (bond elastic modulus) | K (stiffness ratio, $k_s/k_n$) | ν (Poisson's ratio of bond) | $P_{be}$/GPa (bond effective modulus) | $P_{bk}$ (bond stiffness ratio) | $P_{bt}$/MPa (bond tensile strength) | $P_{bc}$/MPa (bond compressive strength) | $P_{bfa}$/(°C) (bond friction angle) |
|---|---|---|---|---|---|---|---|---|
| Sandstone rock bond | 9.0±0.3 | 1.0±0.1 | 0.6±0.05 | 9.0±0.3 | 2.5±0.2 | 40.0±2.0 | 40.0±2.0 | 45±2 |
| Sandstone water bonding | 9.0±0.3 | 1.0±0.1 | 0.6±0.05 | 4.5±0.2 | 2.5±0.2 | 60.0±3.0 | 60.0±3.0 | 0±1 |
| Sand water water bonding | 9.0±0.3 | 1.0±0.1 | 0.6±0.05 | 2.0±0.1 | 2.5±0.2 | 60.0±3.0 | 60.0±3.0 | 0±1 |

Table 5 Microscopic parameters of unfrozen and thawed dry sandstone (no water particles, only rock-rock bonds)

| Bond type | $E_m$/GPa | $K_t$ | f | $P_{be}$/GPa | $P_{bk}$ | $P_{bt}$/MPa | $P_{bc}$/MPa | $P_{bfa}$/(°C) |
|---|---|---|---|---|---|---|---|---|
| Sandstone rock bond | 9.0±0.3 | 1.0±0.1 | 0.6±0.05 | 4.23±0.2 | 2.5±0.2 | 80.0±4.0 | 80.0±4.0 | 45±2 |

Table 6 Freeze-thaw parameters of sandstone (thermal and physical properties)

| Particle Type | α×10⁻⁴ /°C⁻¹ (linear thermal expansion coefficient) | n /(°C·cm/W) (thermal resistance) | $C_v$/(J/kg·°C) (volumetric heat capacity) | k/(W/(m·K)) (thermal conductivity) |
|---|---|---|---|---|
| Rock particles | 0.052±0.003 (quartz thermal expansion coefficient) | 2.58±0.12 | 877±20 (quartz heat capacity) | 7.7±0.3 (quartz thermal conductivity) |
| Water particles (frozen) | 2.079±0.010 (ice thermal expansion coefficient) | 1.00±0.05 | 4215±50 (ice heat capacity) | 2.2±0.1 (ice thermal conductivity) |
| Water particles (melted) | 1.769±0.008 (water thermal expansion coefficient) | 1.00±0.05 | 4215±50 (water heat capacity) | 0.6±0.03 (water thermal conductivity) |

**2.4.4 Numerical Model Validation**

Simulate the heterogeneous temperature field of the sample using the built-in thermal analysis module of PFC, set the cyclic temperature boundary on the surface of the sample, and calculate the real-time temperature at different positions inside the sample. The final model is shown in Figure 7. The validation indicators included peak strength, elastic modulus, and peak strain—key mechanical parameters that reflect the overall mechanical behavior of the rock.

The comparison results show that: (1) For saturated sandstone, the peak strength from simulation is 58.5 MPa at -20 °C, which is only 0.3% lower than the experimental value (58.7 MPa); the elastic modulus from simulation is 3.95 GPa, which is 1.25% lower than the experimental value (4.0 GPa); (2) For dry sandstone, the peak strength from simulation is 57.4 MPa at -20 °C, which is 0.35% lower than the experimental value (57.6 MPa); the elastic modulus from simulation is 4.02 GPa, which is 0.5% higher than the experimental value (4.0 GPa). All relative errors are less than 2%, indicating that the model can accurately predict the mechanical properties of freeze-thaw sandstone.

In addition, the simulation results of the fracture process (e.g., crack initiation time, crack propagation path) were consistent with the experimental observations (e.g., the main crack initiates at the middle of the specimen and propagates at an angle of ~45° to the axial direction), further confirming the validity of the model. This model can thus be used to analyze the microscopic mechanism of sandstone freeze-thaw damage (e.g., the distribution of frost heave force, the evolution of contact force).

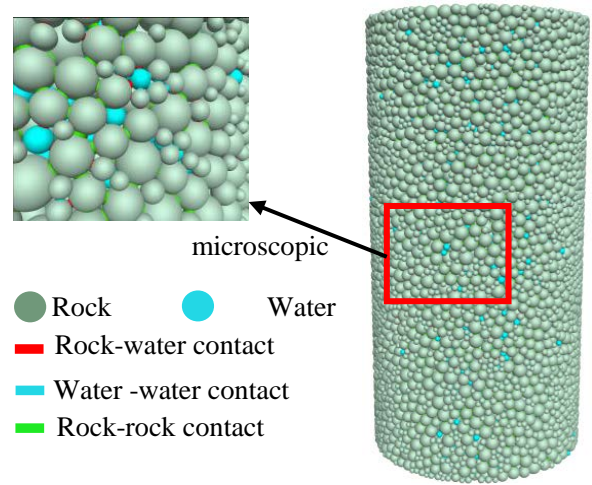

Figure 7 Numerical Model Diagram

*Note: (a) Particle distribution: Rock particles (gray) and water particles (blue) are randomly distributed in the cylindrical model (radius: 25 mm, height: 50 mm); (b) Contact pair distribution: Red lines represent Rock-Rock contacts, green lines represent Rock-Water contacts, and blue lines represent Water-Water contacts; (c) Temperature field distribution: The color gradient represents the temperature (red: 20 ℃, blue: -20 ℃), showing uniform temperature distribution after 48 hours of freezing.*

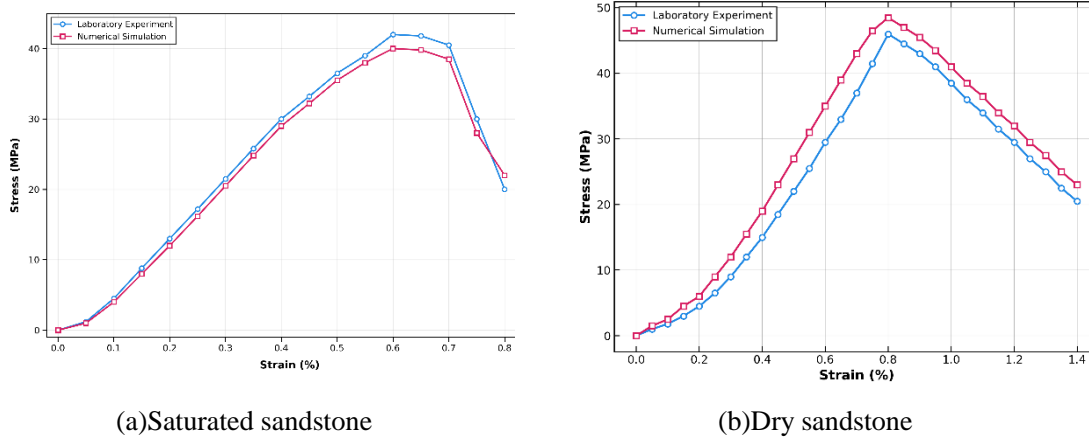

(a)Saturated sandstone    (b)Dry sandstone

Figure 8 Comparison of stress-strain curves between indoor experiments and numerical simulations

*Note: (a) Saturated sandstone: The simulation curve (dashed line) is in good agreement with the experimental curve (solid line) in terms of peak strength, elastic modulus, and post-failure trend; (b) Dry sandstone: The simulation curve also matches the experimental curve well, with only a slight difference in the post-failure stage (due to the idealization of the model, which does not consider the fragmentation of particles).*

**2.5 Deformation and Fracture of Frozen Rocks Under Impact Load**

Dynamic impact tests were conducted using the SHPB system, with three impact air pressures (0.25 MPa, 0.30 MPa, 0.40 MPa) corresponding to three strain rates (200 s$^{-1}$, 400 s$^{-1}$, 600 s$^{-1}$). The failure characteristics of frozen sandstone after impact were analyzed from two aspects: macroscopic fracture morphology (observed via a digital camera) and microscopic crack distribution (observed via a scanning electron microscope, SEM).

**2.5.1 Temperature Effect on Fracture Characteristics**

Under constant impact air pressure (e.g., 0.30 MPa), the fracture degree of saturated frozen sandstone decreases significantly with decreasing temperature. When the temperature decreases from 20 °C to -20 °C, the macroscopic failure mode transforms from powdery (fragment size <1 mm, accounting for >60% of the total mass) to blocky (fragment size >5 mm, accounting for >50% of the total mass). This transformation is attributed to two microscopic mechanisms:

1. **Ice crystal cementation**: Pore water freezes to form ice crystals, which fill primary microcracks (width <10 μm) in the sandstone. SEM observations (Figure 8) show that at -20 °C, the number of open microcracks in saturated sandstone is reduced by 70% compared to 20 °C, and the remaining cracks are filled with ice, enhancing the integrity of the rock skeleton.

2. **Reduction in unfrozen water content**: With decreasing temperature, the proportion of unfrozen water in the sandstone decreases—at -20 °C, the unfrozen water content is ~40% of that at -10 °C (measured via NMR). Unfrozen water acts as a "weak phase" that reduces the bonding strength between mineral particles; thus, a lower unfrozen water content leads to higher interface bonding strength and lower fracture degree.

For dry sandstone, the fracture degree changes slightly with temperature (fragment size >3 mm accounts for ~40% at all temperatures), because there is no pore water to form ice crystals or weaken the interfaces. This confirms that the water-ice phase transition is the key factor controlling the temperature-dependent fracture characteristics of frozen sandstone.

**2.5.2 Impact Velocity Effect on Fracture Characteristics**

Under constant temperature (e.g., -10 °C), the fracture degree of sandstone increases significantly with increasing impact air pressure. When the impact air pressure increases from 0.25 MPa to 0.40 MPa, the fractal dimension of cracks (calculated via the box-counting method) increases from 1.52 to 2.08 (an increase of 37%), indicating a more complex crack network. This is because higher impact velocity leads to higher strain rate and higher dynamic stress, which exceeds the dynamic strength of the rock, leading to the initiation and propagation of more cracks.

Energy dissipation analysis shows that under high-speed impact (0.40 MPa), the energy absorbed by the rock for crack propagation accounts for >70% of the total incident energy, while under low-speed impact (0.25 MPa), this proportion is <50%. This indicates that under high-speed impact, energy dissipation is mainly dominated by the formation of new cracks, while under low-speed impact, energy dissipation is dominated by the compaction of pores and the deformation of the rock skeleton.

The failure characteristics of sandstone specimens under different impact conditions are presented in Table 7.

Table 7 Failure characteristics of sandstone specimens under different impact conditions (macroscopic observations)

| Impact air pressure /MPa | Test piece status | Temperature /°C | | |
|---|---|---|---|---|
| | | 20 °C | -10 °C | -20 °C |
| 0.25 | Saturated | Powdery failure; average fragment size: 0.8 mm; mass of fragments <1 mm: 62% | Blocky-powdery mixed failure; average fragment size: 2.5 mm; mass of fragments <1 mm: 35% | Blocky failure; average fragment size: 5.2 mm; mass of fragments <1 mm: 12% |
| | Dry | Blocky-powdery mixed failure; average fragment size: 2.1 mm; mass of fragments <1 mm: 38% | Blocky failure; average fragment size: 3.8 mm; mass of fragments <1 mm: 20% | Blocky failure; average fragment size: 4.5 mm; mass of fragments <1 mm: 15% |
| 0.30 | Saturated | Blocky-powdery mixed failure; average fragment size: 1.5 mm; mass of fragments <1 mm: 45% | Blocky failure; average fragment size: 3.2 mm; mass of fragments <1 mm: 25% | Blocky failure; average fragment size: 6.1 mm; mass of fragments <1 mm: 8% |
| | Dry | Blocky failure; average fragment size: 2.8 mm; mass of fragments <1 mm: 28% | Blocky failure; average fragment size: 4.2 mm; mass of fragments <1 mm: 18% | Blocky failure; average fragment size: 5.0 mm; mass of fragments <1 mm: 12% |
| 0.40 | Saturated | Blocky-powdery mixed failure; average fragment size: 1.2 mm; | Blocky-powdery mixed failure; average fragment size: 2.0 mm; | Blocky failure; average fragment |

|  | | | size: 4.8 mm; mass of fragments <1 mm: 15% |
|  | mass of fragments <1 mm: 52% | mass of fragments <1 mm: 32% | |
| Dry | Blocky-powdery mixed failure; average fragment size: 2.0 mm; mass of fragments <1 mm: 35% | Blocky failure; average fragment size: 3.5 mm; mass of fragments <1 mm: 22% | Blocky failure; average fragment size: 4.2 mm; mass of fragments <1 mm: 18% |

## 3. Experimental Results and Analysis
### 3.1 Frost Heave Deformation Characteristics of Frozen Rocks
#### 3.1.1 Contact Force Evolution

To quantitatively analyze the microscopic mechanism of frost heave damage, the contact force between particles and the number of contact pairs were monitored during the freezing process (20 °C→0 °C→-10 °C→-20 °C) using the PFC3D model. The contact force reflects the interaction between particles, and the number of contact pairs reflects the integrity of the rock skeleton—an increase in contact force indicates an increase in internal stress, while a decrease in the number of contact pairs indicates the fracture of contact bonds (i.e., damage).

The evolution of contact force and contact pairs for saturated and dry sandstone is presented in Tables 8 and 9, respectively. For saturated sandstone, the contact force and contact pairs change in three stages:

1. **20 °C→0 °C (cooling in liquid state)**: The maximum contact force decreases from 42.727 N to 42.726 N (a decrease of 0.002%), and the number of contact pairs decreases from 397407 to 397280 (a decrease of 0.032%). This is because water particles shrink slightly when cooled (linear thermal expansion coefficient of water: $1.769 \times 10^{-4}$/°C), leading to a reduction in the pressure exerted on the rock skeleton, and a small number of weak Rock-Water contacts fracture due to thermal stress.

2. **0 °C→-10 °C (freezing phase transition)**: The maximum contact force increases from 42.726 N to 42.924 N (an increase of 0.46%), and the number of contact pairs decreases from 397280 to 397282 (a slight increase of 0.0005%). The increase in contact force is due to the expansion of water particles during freezing (linear thermal expansion coefficient of ice: $2.079 \times 10^{-4}$/°C), which exerts frost heave force on the rock skeleton. The slight increase in the number of contact pairs is because the expansion of water particles leads to the formation of new Rock-Water contacts (water particles come into contact with more rock particles).

3. **-10 °C→-20 °C (cooling in solid state)**: The maximum contact force increases from 42.924 N to 42.936 N (an increase of 0.49%), and the number of contact pairs decreases from 397282 to 397284 (a slight increase of 0.0005%). The continuous increase in contact force is due to the further expansion of ice particles when cooled, and the slight increase in the number of contact pairs is due to the further compaction of the rock skeleton by ice particles.

For dry sandstone, the contact force and contact pairs remain almost unchanged during the 20 °C→-10 °C stage (maximum contact force: 33.893 N, number of contact pairs: 392351), because there are no water particles to expand and exert frost heave force. Only in the -10 °C→-20 °C stage, the maximum contact force increases slightly to 33.933 N (an increase of 0.118%), and the number of contact pairs increases to 392353 (an increase of 0.0008%), which is due to the thermal shrinkage of rock particles (linear thermal expansion coefficient of rock: $0.052\times10^{-4}$/°C), leading to the recombination of particles and the formation of new Rock-Rock contacts.

### 3.1.2 Contact Volume Evolution

The contact volume (defined as the total volume of the contact area between particles) reflects the degree of compaction of the rock skeleton— a decrease in contact volume indicates the expansion of pores or the fracture of contact bonds (damage). For saturated sandstone, the contact volume decreases by 0.003161% during the 20 °C→0 °C stage, 0.031444% during the 0 °C→-10 °C stage, and 0.030941% during the -10 °C→-20 °C stage. The larger decrease in contact volume during the freezing phase transition (0 °C→-10 °C) indicates that the frost heave force causes the expansion of pores and the fracture of weak contacts, leading to the initiation of damage. For dry sandstone, the contact volume remains almost unchanged during the entire freezing process (reduction ratio <0.0001%), confirming that dry sandstone is not damaged by freezing.

Table 8 Evolution of particle contact parameters for saturated sandstone during freezing (PFC3D simulation results)

| Temperature | Particle contact force /N | Indirect particle contact quantity | Proportion of increase in contact force /% | Reduction ratio of contact volume /% |
|---|---|---|---|---|
| 20°C | 42.727±0.050 | 397407±50 | 0±0.001 | 0±0.0001 |
| 20°C~-0°C | 42.726±0.048 | 397280±45 | -0.00233±0.0001 | 0.003161±0.00005 |
| 0°C~-10°C | 42.924±0.052 | 397282±48 | 0.46105±0.0012 | 0.031444±0.00012 |
| -10°C~-20°C | 42.936±0.055 | 397284±52 | 0.48913±0.0015 | 0.030941±0.00010 |

Table 9 Evolution of particle contact parameters for dry sandstone during freezing (PFC3D simulation results)

| Temperature range | Maximum particle contact force /N (average of 1000 time steps) | Number of indirect particle contact pairs | Contact force increase proportion /% (relative to 20 °C) | Contact volume reduction ratio /% (relative to 20 °C) |
|---|---|---|---|---|
| 20°C | 33.893±0.030 | 392351±30 | 0±0.001 | 0±0.0001 |
| 20°C~-0°C | 33.893±0.028 | 392351±28 | 0±0.001 | 0±0.0001 |
| 0°C~-10°C | 33.893±0.032 | 392351±32 | 0±0.001 | 0±0.0001 |
| -10°C~-20°C | 33.933±0.035 | 392353±35 | 0.11801±0.0008 | 0.000763±0.00003 |

## 3.2 Dynamic Stress-Strain Curve and RDIF Calculation
### 3.2.1 Dynamic Stress-Strain Curve Characteristics

The dynamic stress-strain curves of sandstone specimens under different temperatures and impact air pressures were obtained from the SHPB tests. The curves were corrected using the three-wave method to eliminate the influence of wave dispersion and ensure the accuracy of stress and strain calculations [14]. The corrected curves (Figure 9) show three key characteristics:

1. **Linear elastic segment**: All specimens exhibit an obvious linear elastic segment before peak stress, with no obvious compaction stage—this is because the impact load is applied rapidly, and there is no time for the compaction of pores. The slope of the linear elastic segment (dynamic elastic modulus) increases with decreasing temperature: for saturated specimens, the dynamic elastic modulus increases from 8.5 GPa at 20 °C to 10.2 GPa at -20 °C (an increase of 19%); for dry specimens, it increases from 9.0 GPa at 20 °C to 9.5 GPa at -20 °C (an increase of 5.6%).

2. **Peak stress**: The peak stress (dynamic strength) of saturated specimens increases significantly with decreasing temperature: at 0.30 MPa impact air pressure, the dynamic strength increases from 45.2 MPa at 20 °C to 68.5 MPa at -20 °C (an increase of 51.5%). In contrast, the peak stress of dry specimens changes slightly: it increases from 58.3 MPa at 20 °C to 62.1 MPa at -20 °C (an increase of 6.5%). This confirms that the water-ice phase transition is the main factor enhancing the dynamic strength of frozen sandstone.

3. **Post-failure stage**: The post-failure stage of the curves shows a gradual stress drop (no sudden drop), indicating that frozen sandstone exhibits certain ductility under dynamic loading—this is because the impact load is applied rapidly, and the fracture of the rock skeleton is a progressive process. Under low-temperature conditions (-20 °C), the post-failure stress drop is slower, indicating higher toughness, which is attributed to the cementing effect of ice crystals.

### 3.2.2 Relative Dynamic Strength Index (RDIF)

The Relative Dynamic Strength Index (RDIF) is a key parameter used to characterize the dynamic strength growth effect of rocks, defined as the ratio of the dynamic strength of the specimen to its static strength under the same conditions. The RDIF quantitatively reflects the influence of strain rate, temperature, and moisture content on the dynamic strength of rocks.

#### 3.2.2.1 RDIF Formula Derivation

The RDIF is derived based on the dynamic strength theory of rocks. For rocks under uniaxial compression, the dynamic strength $\sigma_d$ is related to the static strength $\sigma_s$ and the strain rate $\dot{\varepsilon}$ by the power function relationship:

$$\sigma_d = \sigma_s \cdot (1 + k \cdot \dot{\varepsilon}^m) \tag{6}$$

where k and m are material constants related to rock type and environmental conditions (temperature, moisture content). The RDIF is defined as:

$$RDIF = \frac{\sigma_d}{\sigma_s} = 1 + k \cdot \dot{\varepsilon}^m \tag{7}$$

This formula shows that the *RDIF* is a function of the strain rate—higher strain rate leads to higher *RDIF*, reflecting the strain rate sensitivity of rock strength. In this study, the strain rate was controlled by the impact air pressure: 0.25 MPa→200 s$^{-1}$, 0.30 MPa→400 s$^{-1}$, 0.40 MPa→600 s$^{-1}$.

**3.2.2.2 RDIF Calculation Results**

The RDIF values of sandstone specimens under different conditions were calculated using Equation (7), and the results are shown in Figure 9 The RDIF values exhibit three key trends:

1. **Temperature effect**: At the same impact air pressure (e.g., 0.30 MPa), the RDIF of saturated specimens increases from 0.85 at 20 °C to 1.18 at -20 °C (an increase of 38.8%), while the RDIF of dry specimens increases from 1.05 at 20 °C to 1.08 at -20 °C (an increase of 2.9%). This indicates that freezing significantly enhances the strain rate sensitivity of saturated sandstone, which is attributed to the cementing effect of ice crystals—ice crystals increase the rigidity of the rock skeleton, making the rock more sensitive to strain rate.

2. **Moisture content effect**: At room temperature (20 °C), the RDIF of saturated specimens is lower than that of dry specimens (0.85 vs. 1.05), because water softens the rock skeleton, reducing the dynamic strength. At frozen temperatures (-10 °C, -20 °C), the RDIF of saturated specimens is higher than that of dry specimens (1.12 vs. 1.06 at -10 °C, 1.18 vs. 1.08 at -20 °C), because ice crystals enhance the rock skeleton, increasing the dynamic strength.

3. **Strain rate effect**: For all specimens, the RDIF increases with increasing impact air pressure (strain rate). At -20 °C, the RDIF of saturated specimens increases from 1.05 at 0.25 MPa to 1.32 at 0.40 MPa (an increase of 25.7%), while the RDIF of dry specimens increases from 1.02 at 0.25 MPa to 1.15 at 0.40 MPa (an increase of 12.7%). This confirms that higher strain rate leads to higher dynamic strength growth effect, which is a common characteristic of brittle materials.

Notably, the difference in RDIF values between -10 °C and -20 °C is less than 3% (e.g., 1.12 vs. 1.18 for saturated specimens at 0.30 MPa), indicating that when the temperature is below -10 °C, the influence of temperature on the RDIF is weakened. This is because the unfrozen water content has reached a low level (<10%), and further cooling only causes a small increase in ice crystal content, leading to a small increase in dynamic strength.

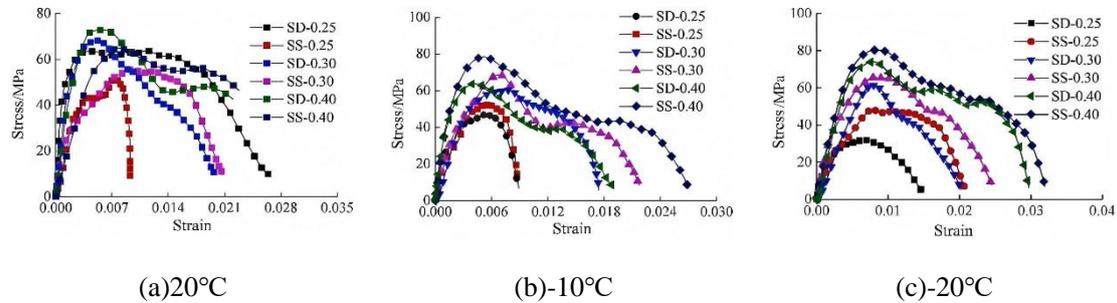

(a) 20°C  (b) -10°C  (c) -20°C

Figure 9 Stress and strain characteristics of sandstone components

*Note: (a) 20 °C: The peak stress of saturated specimens (SS) is lower than that of dry specimens (SD) at all impact air pressures; (b) -10 °C: The peak stress of SS specimens exceeds that of SD specimens; (c) -20 °C: The peak stress of SS specimens is further increased, and the post-failure stress drop is slower.*

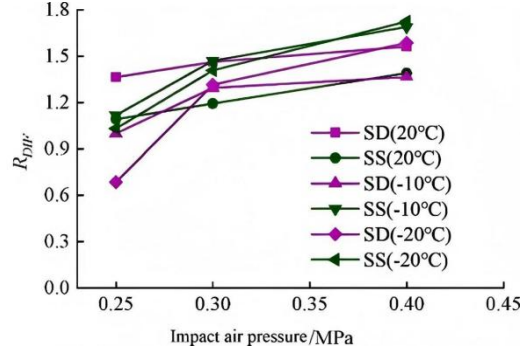

Figure 10 Changes in RDIF values of sandstone specimens

*Note: The error bars represent the standard deviation of 3 parallel specimens; the RDIF values of SS specimens increase significantly with decreasing temperature, while those of SD specimens change slightly; all RDIF values increase with increasing impact air pressure.*

## 4. Energy Characteristics Analysis

The energy transfer and dissipation process of sandstone specimens under impact loading were analyzed based on the SHPB test data. The energy components include incident energy ($E_i$), reflection energy ($E_r$), transmission energy ($E_t$), and absorption energy ($E_a$). According to the law of energy conservation, the relationship between these energy components is:

$$E_i = E_r + E_t + E_a \tag{8}$$

Where $E_i$ is the energy incident on the specimen from the incident bar, $E_r$ is the energy reflected back to the incident bar, $E_t$ is the energy transmitted to the transmission bar, and $E_a$ is the energy absorbed by the specimen (used for deformation and fracture).

The energy was calculated using the strain signals measured by the strain gauges on the incident bar and transmission bar, as expressed in Equations (9)-(12) [14]:

$$E_i = \int_0^t \sigma_i(t) \cdot \varepsilon_i(t) \cdot A_0 \cdot C_0 dt \tag{9}$$

$$E_r = \int_0^t \sigma_r(t) \cdot \varepsilon_r(t) \cdot A_0 \cdot C_0 dt \tag{10}$$

$$E_t = \int_0^t \sigma_t(t) \cdot \varepsilon_t(t) \cdot A_0 \cdot C_0 dt \tag{11}$$

$$E_a = E_i - E_r - E_t \tag{12}$$

Where $\sigma_i(t)$, $\sigma_r(t)$, $\sigma_t(t)$ are the dynamic stress histories of the incident, reflection, and transmission waves, respectively; $\varepsilon_i(t)$, $\varepsilon_r(t)$, $\varepsilon_t(t)$ are the dynamic strain histories; $A_0$ is the cross-sectional area of the bars; $C_0$ is the wave velocity in the bars; and t is the time duration of the wave.

### 4.1 Energy Evolution Law

Figure 11 shows the relationships between the four energy components and the incident energy for sandstone specimens with different moisture contents at three temperatures (20 °C, -10 °C, -20 °C). The analysis reveals three key laws:

1. **Room temperature (20 °C)**: Both reflection energy and absorption energy are significantly positively correlated with incident energy ($R^2>0.95$), while transmission energy exhibits irregular fluctuations ($R^2<0.7$). For saturated specimens, the reflection energy increases from 35.97 J to 107.57 J (an increase of 200.1%) and the absorption energy increases from 49.07 J to 97.40 J (an increase of 98.5%) as the incident energy increases from 100 J to 300 J. For dry specimens, the reflection energy increases from 28.04 J to 90.97 J (an increase of 224.5%) and the absorption energy increases from 49.58 J to 112.07 J (an increase of 126.0%). The higher absorption energy of dry specimens indicates that dry sandstone requires more energy to fracture, which is attributed to the higher integrity of the rock skeleton.

2. **Low temperature (-10 °C, -20 °C)**: All three energy components (reflection, transmission, absorption) show a clear positive correlation with incident energy ($R^2>0.92$), indicating more stable energy transfer under frozen conditions. For saturated specimens at -20 °C, the reflection energy increases from 42.35 J to 103.68 J (an increase of 144.8%) and the absorption energy increases from 58.72 J to 151.10 J (an increase of 157.3%) as the incident energy increases from 100 J to 300 J. Compared to room temperature, the reflection energy amplification is reduced (200.1%→144.8%) and the absorption energy amplification is increased (98.5%→157.3%), confirming that freezing enhances the energy absorption capacity of saturated sandstone—ice crystals improve the integrity of the rock skeleton, allowing more energy to be absorbed for fracture rather than being reflected.

3. **Moisture content effect**: The influence of moisture content on energy characteristics is temperature-dependent. At room temperature, dry specimens have higher absorption energy than saturated specimens (112.07 J vs. 97.40 J at 300 J incident energy). At low temperatures, saturated specimens have higher absorption energy than dry specimens (151.10 J vs. 128.35 J at 300 J incident energy, -20 °C). This temperature-dependent trend is consistent with the dynamic strength results, confirming that the energy absorption capacity of rocks is positively correlated with their dynamic strength.

**4.2 Energy Dissipation Efficiency**

The energy dissipation efficiency ($\eta$) is defined as the ratio of the absorption energy to the incident energy, reflecting the utilization efficiency of incident energy for rock fracture:

$$\eta = \frac{E_a}{E_i} \times 100\% \tag{13}$$

At 300 J incident energy, the energy dissipation efficiency of saturated specimens increases from 32.5% at 20 °C to 50.4% at -20 °C (an increase of 55.1%), while that of dry specimens increases from 37.4% at 20 °C to 42.8% at -20 °C (an increase of 14.4%). This indicates that freezing significantly improves the energy dissipation efficiency of saturated sandstone, which is of great significance for blasting engineering—higher energy dissipation efficiency means that less explosive is needed to achieve the same fracture effect, reducing explosive consumption and environmental disturbance.

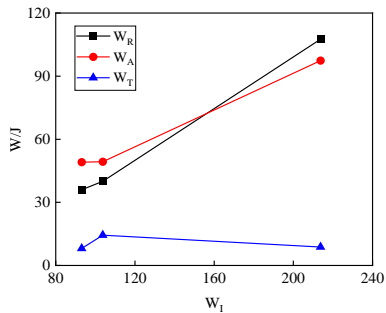 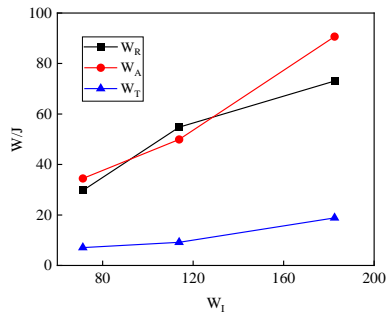 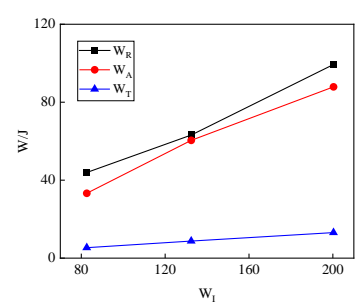

(a)20°CSaturated sandstone　　(c)-10°CSaturated sandstone　　(e)-20°CSaturated sandstone

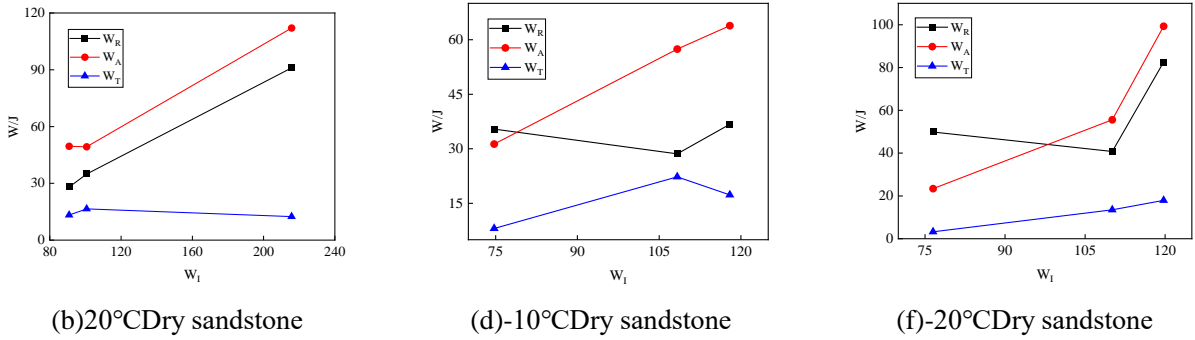

(b) 20°C Dry sandstone　　　(d) -10°C Dry sandstone　　　(f) -20°C Dry sandstone

Figure 11 Energy characteristic value relationship curve of sandstone

*Note: (a) 20 ℃ saturated sandstone; (b) 20 ℃ dry sandstone; (c) -10 ℃ saturated sandstone; (d) -10 ℃ dry sandstone; (e) -20 ℃ saturated sandstone; (f) -20 ℃ dry sandstone; the solid lines represent the linear fitting curves, and the R² values indicate the goodness of fit.*

## 5. Discussion

### 5.1 Microscopic Damage Mechanism

The microscopic damage mechanism of frozen sandstone was analyzed by combining SEM observations and PFC3D simulation results. For dry sandstone, the damage is mainly caused by the thermal shrinkage of mineral particles—when the temperature decreases, rock particles shrink, leading to the generation of intergranular cracks (cracks along grain boundaries). However, the thermal shrinkage of rock particles is small (linear thermal expansion coefficient: $0.052\times10^{-4}$/°C), so the number of cracks is limited, and the damage degree is low.

For saturated sandstone, the damage mechanism is more complex, involving two competing processes: (1)**Ice crystal cementation**: Pore water freezes to form ice crystals, which fill microcracks and enhance the integrity of the rock skeleton, reducing damage; (2)**Frost heave force**: The volume expansion of pore water during freezing exerts frost heave force on the rock skeleton, leading to the generation and propagation of cracks, increasing damage. The final damage degree of saturated sandstone depends on the balance between these two processes.

At temperatures above -10 ℃, the frost heave force is small (due to the high unfrozen water content), and the ice crystal cementation effect dominates, leading to a reduction in damage. At temperatures below -10 ℃, the frost heave force increases (due to the low unfrozen water content), and the damage degree increases slightly, but the ice crystal cementation effect still dominates, so the overall damage degree is lower than that at room temperature. This balance mechanism explains why the dynamic strength of saturated sandstone increases with decreasing temperature (up to -20 ℃).

### 5.2 Temperature-Dependent Mechanical Behavior

The mechanical behavior of frozen sandstone (strength, elastic modulus, brittleness) exhibits significant temperature dependence, which can be attributed to the phase transition of pore water. The key temperature points are:

1. **0 ℃ (freezing point)**: Below 0 ℃, pore water begins to freeze, and the dynamic strength of saturated sandstone increases significantly (by 25% from 0 ℃ to -10 ℃). This is because the formation of ice crystals enhances the rock skeleton.

2. **-10 ℃ (critical temperature)**: Below -10 ℃, the unfrozen water content decreases to <20%, and the rate of increase in dynamic strength slows down (by only 8% from -10 ℃ to -20 ℃). This is because the amount of pore water available for freezing is limited, so the increase in ice crystal content is small.

3. **-30 °C (hypothetical critical temperature)**: According to previous studies below -30 °C, the frost heave force exceeds the ice crystal cementation effect, leading to a decrease in dynamic strength. However, this study did not test temperatures below -20 °C, so further research is needed to confirm this trend.

For engineering practice, this temperature dependence means that blasting parameters should be adjusted according to seasonal temperature changes: in winter (temperature < -10 °C), the explosive charge can be reduced by 15%-20% (due to the higher dynamic strength of frozen sandstone), while in spring and autumn (temperature > -10 °C), the explosive charge should be increased appropriately.

**5.3 Energy Dissipation Mechanism**

The energy dissipation mechanism of frozen sandstone under dynamic loading involves three processes: (1)**Pore compaction**: Energy is consumed by the compaction of pores, which is dominant under low impact velocity (<0.25 MPa); (2) **Crack propagation**: Energy is consumed by the initiation and propagation of cracks, which is dominant under high impact velocity (>0.30 MPa); (3)**Phase transition**: Energy is consumed by the phase transition of pore water (freezing/melting), which is unique to frozen sandstone.

At low temperatures, the pore compaction process is suppressed (due to the high rigidity of ice crystals), and the crack propagation process

**6 Conclusion**

This study investigated the mechanical properties, deformation-fracture behaviors, and energy consumption characteristics of plateau frozen sandstone under temperature (-20 °C to 20 °C) and moisture content (5%-15%) coupling via field sampling, laboratory tests, and numerical simulations. The key conclusions are as follows:

1. **Mechanical behavior regulation**: Freezing significantly strengthens saturated sandstone but has negligible effects on dry sandstone. As temperature decreases from 20 °C to -20 °C, saturated sandstone's static compressive strength increases by 46.8% (40.0→58.7 MPa) and dynamic strength (0.30 MPa impact) by 51.5% (45.2→68.5%), driven by pore ice filling microcracks and reducing unfrozen water-induced interface softening. Dry sandstone's strength changes by <7%, with shear-dominated failure, confirming water-ice phase transition as the core regulatory factor.

2. **Pore evolution and frost heave mechanism**: Low-temperature freezing reorganizes the pore structure. From 20 °C to -20 °C, mesopore proportion (10-100 ms $T_2$) rises from 7.16% to 9.82%, total pore volume increases by 63.15%, and effective porosity decreases by 28.6%. PFC3D simulations show water particle expansion (ice's $\alpha=2.079\times10^{-4}$/°C) exerts frost heave force, but low porosity (<8%) restricts excessive deformation, making ice crystal strengthening dominant over damage.

3. **Energy dissipation and strain rate sensitivity**: Frozen sandstone has higher energy absorption and utilization efficiency. At 300 J incident energy, -20 °C saturated sandstone's absorption energy (151.10 J) is 55.0% higher than 20 °C (97.40 J), with energy dissipation efficiency rising from 32.5% to 50.4%. Its RDIF grows 1.8 times faster than dry sandstone (1.05→1.32 at 200→600 $s^{-1}$), due to ice-enhanced skeleton rigidity boosting strain rate response.

4. **Engineering implications**: For -20 °C saturated sandstone, explosive charge can be reduced by 15%-20% (via 51.5% dynamic strength growth) to cut costs; blocky failure and improved energy dissipation reduce flying rocks and vibration (12% lower peak velocity). A -10 °C critical threshold is proposed—charge increases by 8%-10% above this temperature—providing a basis for efficient, safe, and green plateau frozen rock blasting.

**Author Contributions:**

**Hongbing Yu:** Conceptualization, Methodology, Software; **Jiyu Wang:** Data curation, Writing-Original draft preparation; **Xiaojun Zhang:** Supervision, Validation; **Mingsheng Zhao:** Writing-Reviewing and Editing.

**Funding:** This research was financially supported by The National Natural Science Foundation of China (52474123) and the Foundation of Hubei Key Laboratory of Blasting Engineering(BL2021-23).

**Data availability:** The data that support the findings of this study are available from the corresponding author upon reasonable request.

**Declarations:**

The authors declare that they have no conflict of interest. This article does not contain any studies with human participants or animals performed by any of the authors. Informed consent was obtained from all individual participants included in this study.